\documentclass[pra,twocolumn,showpacs,preprintnumbers,amsmath,amssymb]{revtex4}

\usepackage{graphicx}
\usepackage{dcolumn}
\usepackage{bm}

\begin{document}

\title{Entangling the lattice clock: Towards Heisenberg-limited timekeeping}
\author{Jonathan D. Weinstein}
\affiliation{Department of Physics, University of Nevada, Reno NV 89557}
\author{Kyle Beloy}
\affiliation{Department of Physics, University of Nevada, Reno NV 89557}
\author{Andrei Derevianko}
\affiliation{Department of Physics, University of Nevada, Reno NV 89557}

\date{\today}

\begin{abstract}
We present a scheme for entangling 
the atoms of an optical lattice to reduce the quantum projection noise of a clock measurement. The divalent clock atoms 
are held in a lattice at a ``magic'' wavelength that does not perturb the clock frequency -- to maintain clock accuracy -- while an open-shell $J=1/2$ ``head'' atom is coherently transported between lattice sites via the lattice polarization.  This polarization-dependent ``Archimedes' screw'' transport at magic wavelength takes advantage of the vanishing vector polarizability of the scalar, $J=0$, clock states of bosonic isotopes of divalent atoms.  The on-site interactions between the clock atoms and the head atom are used to engineer entanglement and for clock readout.
\end{abstract}





\maketitle



Quantum entanglement is a crucial resource in quantum computing 
and has the potential to improve precision measurements \cite{NielsenChuang00, Preskill00PrecisionMeasurement}.
Here we propose a scheme for entangling an ensemble of several thousands of neutral atoms,
with the specific goal of demonstrating the power of entanglement for measuring time.

Measuring time  with atoms relies on the fact that the quantum-mechanical probability
of making a transition between two clock levels depends on the detuning $\Delta \omega$
of the  probe
field $\omega$ from the atomic transition frequency $\omega_0$. 
By measuring the probability as a function of $\omega$, one can infer if the two frequencies
are equal and thereby ``lock'' a local oscillator to the atomic transition.
Counting the number of oscillations of the local oscillator tells time.
The precision of measuring  $\Delta \omega$ 
is limited by the quantum projection noise \cite{PhysRevA.47.3554}.  For a measurement of $N$ unentangled atoms the resulting signal-to-noise of $\Delta \omega$ scales as $\sqrt{N}$:  the standard quantum limit (SQL). The  use of entanglement holds the promise of improving clock precision to the Heisenberg limit, with  signal-to-noise scaling as $N$.

Measurements with uncertainty below the SQL
may be achieved with squeezed atomic states
\cite{Mandel98QNDSqueezing,VuleticQNDclock,YeQNDproposal}.
While this technique can address large number samples of atoms, squeezing experiments have
attained signal-to-noise ratios far from the Heisenberg limit.
In other work, clock measurements at the Heisenberg limit have been demonstrated for small numbers of entangled ions in traps \cite{Wineland04}.
Those experiments created maximally entangled states to achieve measurements at the Heisenberg limit via a Ramsey-type measurement protocol:  a generalized $\pi/2$ pulse creates a Greenberger-Horne-Zeilinger (GHZ) state, the atoms undergo free evolution, and a final generalized $\pi/2$ pulse is used for readout.
However, scaling ion traps up to entangle increasingly larger number of ions currently remains a work in progress.

Using atom-atom interactions to engineer entanglement between neutral atoms trapped in a lattice may offer the best of both worlds: maximally entangled large-number samples.  Previous proposals have noted the virtues of using alkaline-earth-like atoms in lattices for quantum information and quantum computing \cite{DerCan04,HayJulDeu07,GorReyDal09,DalBoyYe08,ShiKatYam09}; here we focus on entangling atoms for improving the atomic clock.

\begin{figure}[h]
\begin{center}
\includegraphics*[scale=0.58]{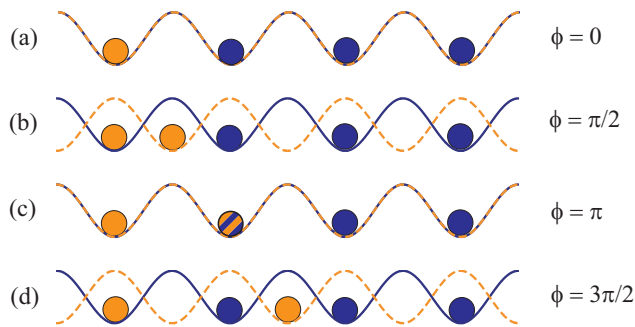}
\end{center}
\caption{(Color online) Schematic of the entanglement process. (a)
With $\phi=0$ a single head atom (orange circle) and several clock atoms
(blue circles) are trapped in the minima of a 1-D optical lattice, with one or fewer atoms per site.
Due to an intensity differential of the underlying lattices, the clock atoms
couple strongly to the $\sigma_+$ lattice (solid blue line). The head
atom is placed in a superposition of atomic states: one which couples strongly to the $\sigma_+$ lattice and one which couples strongly to the $\sigma_-$ lattice (dashed
orange line). (b) As $\phi$ is increased, the latter state spatially
separates and is transported along the lattice. (c)
This portion of the head atom is then brought into contact with a clock
atom to entangle the two atoms. (d) The head atom is transported
further to obtain entanglement with the remaining clock atoms in a
similar manner.} \label{Fig:transport}
\end{figure}

In optical lattice clocks, millions of divalent atoms (such as Sr or Yb) are trapped in an optical lattice (register) operating
at a ``magic'' wavelength $\lambda_m$. At this wavelength, both clock levels, $^1S_0$ and  $^3\!P_0$,
are shifted by the lattice lasers equally, so that the clock frequency remains unperturbed~\cite{KatoriMagic03}.
These clocks have demonstrated long coherence times and have already realized great improvements in both accuracy and precision over the current primary frequency  standard~\cite{A.D.Ludlow03282008,T.Rosenband03282008}.
Notably  
the signal-to-noise ratio of current-generation clocks is approaching the SQL \cite{A.D.Ludlow03282008}, so an ``entangled clockwork'' may be of  practical benefit.

We would like to entangle a string of strontium clock atoms. Each Sr atom occupies an individual lattice site
 as shown in Fig.~\ref{Fig:transport}.
To maintain the clock accuracy we require all lattice fields be at the magic wavelength of strontium.
Entangling the atoms using short-range atom-atom interactions requires the transport of atoms between lattice sites.
While coherent transport in optical lattices has been demonstrated
before using lattice polarization~\cite{PhysRevLett.82.1975,porto07,PhysRevLett.91.010407,Weiss04transport}, such techniques would not work for the clock states of bosonic isotopes, which have a scalar ($F=J=I=0$) nature: the
optical potential 
does not depend on the polarization (be it circular or linear) of the lattice lasers.
Instead, we use a single 
$J=1/2$ ``head'' atom, which is transported from site to site using the optical polarization \cite{PhysRevA.70.012306}.

{\em Lattice ---}  %
Consider the superposition of two spatially-displaced standing waves
of opposite circular polarization $(\hat{\bm{\sigma}}_{\pm})$.
The resulting E-field reads
\begin{equation}
\mathbf{E}(z)=E_+\hat{\bm{\sigma}}_+\mathrm{cos}(\frac{2\pi}{\lambda_m} z)+
E_-\hat{\bm{\sigma}}_-\mathrm{cos}(\frac{2\pi}{\lambda_m} z-\varphi),
\label{Eq:EfieldLattice}
\end{equation}
where winding the phase $\varphi$ similarly to the ``Archimedes' screw''controls a relative displacement between the nodes
of the two standing waves. 
The resulting optical potential reads
\begin{equation}
U(z)=U_0^+\mathrm{cos}^2(\frac{2\pi}{\lambda_m} z)+
U_0^-\mathrm{cos}^2(\frac{2\pi}{\lambda_m} z-\varphi).
\end{equation}
For an atom in an $|F,M_F\rangle$ state (with the quantization axis taken to align with the lattice lasers)
\begin{equation}
U_0^\pm=-\left(\frac{E_\pm}{2}\right)^2\left(\alpha^{\mathrm{s}}(\omega_m) \pm
\frac{M_F}{2 F} \, \alpha_{F}^{\mathrm{a}}(\omega_m) \right).
\end{equation}
Here $\alpha^{\mathrm{s}}(\omega_m)$ and $\alpha_{F}^{\mathrm{a}}(\omega_m)$ are
frequency-dependent scalar and vector (axial) polarizabilities. Neglected tensor
contribution is suppressed~\cite{BelDerDzu08Clock} for $J=1/2$ atoms.

The two clock states $|0\rangle=|^1S_0\rangle$ and $|1\rangle=|^3P_0\rangle$ will experience
the same trapping potential; at $\lambda_m$ the two ac polarizabilities are the same.
Note that the vector part of the polarizability is zero for the scalar clock states.
If two states of the head atom $|\uparrow\rangle=|F,M_F\rangle $
and $|\downarrow\rangle=|F',M_F'\rangle $ have different
vector polarizabilities,
they will see different potentials.
For an appropriate choice of lattice parameters, discussed below,
the  $|\downarrow\rangle$ state  couples preferentially to the $\sigma^+$ permanent
sub-lattice, while  $|\uparrow\rangle$ couples preferentially to the moving $\sigma^-$ sub-lattice.
Unfortunately, the commonly employed lin $\angle$ lin transport lattice algorithms \cite{PhysRevLett.82.1975,porto07,PhysRevLett.91.010407,Weiss04transport}
 cannot be employed here directly, as the potential will wash out for the clock atom in the lin $\perp$ lin  configuration.  However, for appropriate choice of lattice intensities (discussed below), the clock atoms remain pinned to the $\sigma^+$ permanent sub-lattice.
%
%
This state-selective transport enables the entanglement of the clock atoms with the head atom state, as shown in Fig.~\ref{Fig:transport} and described below.

{\em Clock protocol ---}
We describe the system as the product of the state
of the clock register and the state of the head atom. For example, a possible
basis state of $N=3$ clock atoms and a head atom in the spin-up state is
$
|\Psi\rangle= |110\rangle |\uparrow\rangle=|6\rangle|\uparrow\rangle.
$
We require two gates: a single-qubit Hadamard gate
$H$ (an analog of a $\pi/2$
pulse) and a two-qubit phase gate. No individual addressing is required.
The phase gate $P_{i}$ involves the ``head''  atom state-selectively transported to overlap with the target
clock atom at position $i$ in the lattice register, %
$P_{i}|...,0_{i},...\rangle|    \uparrow\rangle=+|...,0_{i},...\rangle
|\uparrow\rangle$  and
$P_{i}|...,1_{i},...\rangle|    \uparrow\rangle=-|...,1_{i},...\rangle
|\uparrow\rangle$.

A practical realization of the Hadamard gate involves interaction with a
near-resonant pulse of optical frequency $\omega$ for the clock qubits and a near-resonant pulse of microwave frequency $\omega^{\prime}$ for the head atom. We want to measure
the clock frequency $\omega_{0}$ by tuning the driving frequency $\omega$.
Below we show that, as in the conventional Ramsey-type clock frequency
measurement, the probability of making a  clock transition depends on the
detuning $\Delta\omega=$ $\omega-\omega_{0}$, allowing to zero-in on the clock
frequency. Theoretical analysis is simplified by transforming into a rotating
reference frame; the relevant chain of operators for the compound wavefunction
involves a product of $N$ clock-state rotation operators at frequency $\omega$
and a rotation operator for the head atom at frequency $\omega^{\prime}$.

We start by filling a 1-D lattice with a single head atom and $N$\ clock
atoms,
$
|\Psi_{0}\rangle=|000...\rangle|\downarrow\rangle.
$
Next we apply the Hadamard gates to the head atom and to all the clock atoms:
\[
|\Psi_{1}\rangle=\left(  \frac{1}{\sqrt{2^{N}}}\sum_{p}|p\rangle\right)
\left(  \frac{1}{\sqrt{2}}\left(  |\downarrow\rangle+|\uparrow\rangle\right)
\right)  ,
\]
where $|p\rangle=|0...00\rangle,|0...01\rangle,...,|1...11\rangle$ is the
computational (binary) basis set for the clock register.  In the next step we
use the transport lattice and move the $|\uparrow\rangle$ state of the head
atom along the clock register to perform a collisional phase gate at each
site.

We presume that the  $|0\rangle$ and $|1\rangle$  clock states
will have different scattering lengths for their interaction with
the $|\uparrow\rangle$ state of the head atom, and thus a different mean-field interaction.
The transport lattice will be moved in such a way that the head atom will remain on-site with
each clock atom for a sufficient period of time to produce a relative phase
shift of $\pi$ \cite{PhysRevLett.82.1975}.
This generates the entangled wavefunction%
\[
|\Psi_{2}\rangle=\frac{1}{\sqrt{2^{N+1}}}\left(  \sum_{p}|p\rangle
|\downarrow\rangle+\sum_{p}\left(  -1\right)  ^{k_{p}}|p\rangle|\uparrow
\rangle\right)  .
\]
Here $k_{p}=\sum_{j}p_{j},~p=\sum_{j=0}^{N-1}p_{j}2^{j}$, i.e., it is the
number of raised bits in the binary representation of $p$.
By applying the Hadamard gate to the clock register we obtain the GHZ state%
\[
|\Psi_\mathrm{GHZ}\rangle=\frac{1}{\sqrt{2}}\left(  |000...000\rangle|\downarrow
\rangle+|111...111\rangle|\uparrow\rangle\right)  .
\]
The entire
procedure may be considered as a generalized $\pi/2$ pulse in the space of
maximally-polarized states.

As in traditional Ramsey spectroscopy
we let the GHZ state evolve freely for a time
$T$.
In the rotating reference
frame, a phase $\chi=\left(  N\Delta\omega+\Delta\omega^{\prime
}\right)  T$ is accumulated during the free evolution. Notice that
$\Delta\omega$ is the detuning for the \emph{optical} clock frequency, while
$\Delta\omega^{\prime}$ is the detuning for the \emph{microwave} transition of
the head atom. Therefore $\Delta\omega^{\prime}\ll N\Delta\omega.$
Finally, we repeat the generalized $\pi/2$ pulse, arriving at%
\[
|\Psi_{\mathrm{final}}\rangle=|0...0\rangle\left\{  \cos\frac{\chi}{2}%
~|\downarrow\rangle-i\sin\frac{\chi}{2}~|\uparrow\rangle\right\}  ,
\]
where the $N$-enhanced phase is encoded into the state of the microwave head qubit.

For readout of the clock, the state of the head qubit can be read directly via laser-induced fluorescence.  In the event that the light collection efficiency is insufficient to read out the state with high efficiency \cite{IonClockReadout}, the state of the head atom can be transferred to many alkaline earth atoms (as done in the first stage of the entanglement algorithm) for efficient readout.


{\em  Choice of the ``head'' atom and lattice parameters ---}
We start with a detailed analysis of the transport lattice which dictates the choice of the head atom.
We introduce the fractional intensity misbalance for the two circularly polarized
sublattices of Eq.~(\ref{Eq:EfieldLattice}):  $\delta \equiv (E_+^2-E_-^2)/(E_+^2+E_-^2)$.
It  does not depend on atomic properties and  $|\delta| \le 1$.
For the clock atoms, the minimum depth of the optical potential ---
which occurs at $\varphi = (n+1/2)\pi$, or positions (b) and (d) in Fig. \ref{Fig:transport} ---
is proportional to $|\delta|$.
However, the larger $|\delta|$ is, the weaker the lattice becomes for one of the head atom states.
The fixed value of $\delta = +1/4$ will be used in all the following calculations.

We further introduce a ratio of the
vector and scalar polarizabilities,
\begin{equation}
\rho=
\frac{M}{2F} \,  \frac{  \alpha_{F}^{\mathrm{a}} (\omega_m)}{
\alpha^{\mathrm{s}} (\omega_m)} \, .
\label{Eq:rho}
\end{equation}
%
The ability to translate the $\left|\uparrow \right\rangle$  state while
holding the $\left|\downarrow \right\rangle$  state stationary is determined by this ratio.
For example,
for $\rho_\uparrow=-1$, $U_0^+=0$ and  the atom does not see the stationary $\sigma^+$
lattice. Similarly, for   $\rho_\downarrow=+1$ the atom has no coupling to the moving
$\sigma^-$ lattice.
In general, for a positive value of $\delta$ we must satisfy the criteria
$-1/\delta< \rho_\uparrow <-\delta$ and
$\rho_\downarrow  < 1/\delta$ together with $\rho_\downarrow  >\delta$.

Qualitatively, for the head atom transport one needs  $|\alpha_{F}^{\mathrm{a}} (\omega_m)| \sim
|\alpha^{\mathrm{S}} (\omega_m)|$. It turns out that at the magic wavelengths specific
to divalent atoms (see Ref.~\cite{DerObrDzu09}),
none of the commonly used alkali-metal atoms satisfies this constraint. Alkali atoms have
the $nS_{1/2}$ ground states and the vector polarizability (rank 1 tensor)
arises only due to relativistic effects. Fortunately, atoms with $nP_{1/2}$
ground states (group III) have large vector polarizabilities.  Aluminum
is a suitable choice for the head atom.

$^{27}$Al has a $3p_{1/2}$ ground state.  The nuclear spin of 5/2 gives
rise to  two hyperfine structure levels: $F=3$ and $F=2$, separated
by 1.5 GHz.  Cooling Al has
already been demonstrated \cite{McGGilLee95} with the goal of atomic
nanofabrication. The laser cooling was carried out on the closed
$3p_{3/2}-3d_{5/2}$ transition with the recoil limit of 7.5 $\mu$K.
Once trapped, the atoms can be readily transferred from the
metastable $3p_{3/2}$ cooling state to the ground (head) state.
Lattice-trapped Al was also considered for quantum information
processing~\cite{RavDerBer06} and for a microwave lattice clock (microMagic clock)~\cite{BelDerDzu08Clock}.

\begin{figure}[h]
\begin{center}
\includegraphics*[scale=0.45]{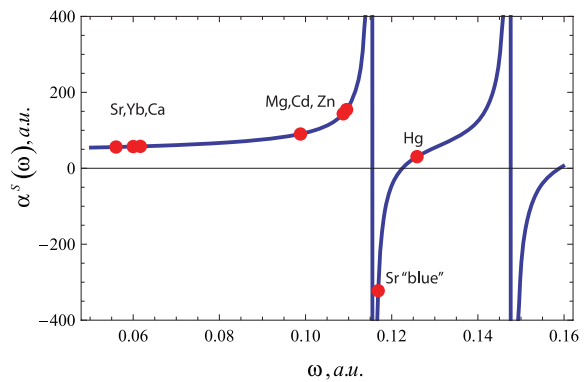}
\end{center}
\caption{ (Color online) Dynamic polarizability of Al as a function of lattice laser frequency. All values are
given in atomic units. Marked points on the plot correspond to magic wavelengths for clock transitions in divalent
atoms
~\cite{DerObrDzu09,TakKatMar09}.
} \label{Fig:polarizAl}
\end{figure}

To evaluate the dynamic polarizabilities for Al,
we employed {\em ab initio} methods of relativistic many-body theory.
To improve on the positions of atomic resonances, for low-lying energy levels we
replaced the {\em ab initio} energies with experimental values. The resulting dynamic
polarizabilities of Al are shown in Fig.~\ref{Fig:polarizAl}.

We focus on the promising case of the Sr ``blue'' magic wavelength~\cite{TakKatMar09} at $\lambda_m^b=389.9$~nm.
Here the clock atoms are confined to minima of laser intensity, reducing photon scattering. Scalar polarizabilities are $\alpha_\mathrm{Sr}^{\mathrm{s}} (\omega_m) \simeq-470~\mathrm{a.u.}$ and $\alpha_\mathrm{Al}^{\mathrm{s}} (\omega_m) \simeq-340~\mathrm{a.u.}$
$^{27}$Al offers many viable $|F,M_F \rangle$ states to implement transport: $\left|\uparrow \right\rangle = |2,2\rangle,|3,-3\rangle,$~or~$|3,-2\rangle$
($\rho \approx -0.84,-1.25,-0.84$, respectively)
and  $\left|\downarrow \right\rangle = |2,-2\rangle,|3,3\rangle,$~or~$|3,2\rangle$  ($\rho \approx 0.84,1.25,0.84$), for example.
We choose $\left|\uparrow \right\rangle = |3,-3\rangle$ and  $\left|\downarrow \right\rangle = |2,-2\rangle$.

Due to $\lambda_m^b$ being  ``blue-detuned'' for both Al and Sr, the atoms will be confined to regions of intensity minima. While the transport lattice provides axial confinement, the radial confinement is provided by two transverse ``blue-detuned'' magic lattices with polarization parallel to the transport lattice's k-vector (to avoid interference with the transport lattice).  The transverse blue lattices create a series of tubes \cite{Bloch05LatticeReview}, each containing a transport lattice.  This provides tight radial confinement and enables many transport lattices to be run in parallel.  We note that to prepare the initial states in the resulting 3-D lattice,  both species can be prepared in the ground vibrational states of individual lattice sites using 3-D Raman sideband cooling \cite{PhysRevLett.84.439,PhysRevLett.85.724}.

The depth $\Delta U$  of the optical potentials is critical, as it determines the rate of unwelcome diffusion out of sites~\cite{DerCan04}.  The depth varies with the displacement phase $\varphi$. For our choice of $| \uparrow \rangle$, the depth 
is weakest when it is on-site with Sr.
At this maximum-overlap position,
$\Delta U (\left|\uparrow \right\rangle)=2 I_L \pi/c |\alpha_\mathrm{Al}^s(\omega_m)| (1+\delta \rho_\uparrow)$ and $\Delta U (\mathrm{clock}) = 2 I_L \pi/c |\alpha_\mathrm{Sr}^s(\omega_m)| $, where $I_L=\frac{c}{8\pi}\left(E_+^2+E_-^2\right)$ and $c$ is the speed of light.
We require $\Delta U > 5 \, E_R$, where $E_R=(2\pi \hbar /\lambda_m)^2/(2M)$ is the recoil energy for an atom of mass $M$. This translates into a minimum intensity of $I_L \sim 20 \, \mathrm{kW/cm}^2$, determined by the lighter Al.

{\em Gate times and decoherence ---}
The number of clock atoms which may be entangled will be limited by
decoherence and phase-gate times.
The main sources of decoherence are anticipated to be inelastic collisions between the clock
atoms and the head atom and light scattering.

The decoherence rate due to the  photon
scattering is
$
\tau_h^{-1}=\eta \frac{8\pi}{3c^4} \, \omega_m^3 \, \left|\alpha^{\mathrm{s}} (\omega_m) \right|^2 I_L
$. Here $\eta\approx 1/2 \sqrt{E_R/\Delta U}$ is a suppression factor~\cite{DerCan04} accounting for atomic wave functions being centered at zero intensity.
We find $\tau_h \approx 10 \, \mathrm{s}$ for Sr and $\tau_h \approx 8\, \mathrm{s}$ for Al.

To estimate the time required for a phase gate operation, we note the interaction energy of two
particles in overlapped ground states of independent 3-D anisotropic harmonic
potentials is given by
$
\delta
E=\frac{2a_\mathrm{scatt}}{\overline{m}}\sqrt{\frac{\hbar}{\pi}}
\prod_{i=xyz}\left(\overline{m\omega}\right)_i^{1/2}
$
where $a_\mathrm{scatt}$ is the scattering length, $\overline{m}$ is
the reduced mass, and $\left(\overline{m\omega}\right)_i$ is defined
analogously to the reduced mass with $\omega_i$ being trap
frequencies.

The axial trap frequencies may be determined from the parameters
above ($I_L,\delta,$ etc.) at overlap. The radial trap
frequencies are determined assuming the transverse lattices are operated at the same intensity as the transport lattice.
Estimating the difference in the excited-state and ground-state scattering lengths to be
$a_\mathrm{scatt}\approx100~\mathrm{a.u.}$ yields an estimated gate
time of
$\tau\approx 20~\mu\mathrm{s}$
and a transport time of $ \tau \approx 10~\mu$s.

\emph{Conclusions ---}
With the pessimistic assumption that a single 
photon scattering event will decohere all the clock atoms, we expect that -- within each 1-D lattice -- one would be able to put $\sim 10^3$ Sr atoms into the maximally entangled GHZ state with high probability.  This would enable a reduction in the projection noise of
lattice clocks.  Moreover, we note that many of the usual requirements for producing highly-entangled states between atoms -- such as single-site addressability, single-site  readout, and unity site occupation -- are absent in this scheme.

Possible improvements would include the use of entangled states less sensitive to photon scattering as a source of decoherence.
%
Moreover, it is possible that  combining these techniques  with more sophisticated  gate operations could lead to the development of a full quantum computer.  In that case, error correction techniques could potentially be used to further increase the number of atoms while maintaining high fidelity.

But even without these improvements, we note that this scheme occupies an interesting ``middle-ground'' of experimental schemes for clock entanglement.  It holds promise for use with larger numbers of atoms than has been demonstrated to date with ion traps.   And while it cannot entangle as large-number samples as are used in spin-squeezing experiments, it may be able to produce greater levels of entanglement.

Unanswered questions remain, such as the value of the Al--Sr scattering length (which determines the gate time) and the rate coefficient for inelastic Al--Sr collisions (which is an additional source of decoherence).  With 3 naturally occurring bosonic isotopes of Sr, it is likely that a favorable scattering length can be found.  The inelastic rate is unknown, but we note that measurements of Al-group atoms in the $^2P_{1/2}$ ground state have observed slow hyperfine relaxation in collisions with $J=0$ noble gas atoms \cite{Weinstein2009InGa}.

\emph{Acknowledgements ---}
We thank V. Vuletic, J. Ye, and A. Gorshkov for useful comments.  This work was supported in part by the US NSF and by the US NASA under Grant/Cooperative Agreement No. NNX07AT65A issued by the Nevada NASA EPSCoR program.

\bibliographystyle{prl_my}

\end{document}